\documentclass[9pt,twocolumn]{revtex4}
\usepackage{amsmath}
\usepackage{eurosym}
\usepackage{color}
\usepackage{graphicx}

\begin{document}

\title{Dark solitons in dual-core waveguides with dispersive coupling}

\author{Yaroslav V. Kartashov$^{1,2}$, Vladimir V. Konotop$^{3}$, and Boris A. Malomed$^{4}$}

\affiliation{$^1$ICFO-Institut de Ciencies Fotoniques, and Universitat Politecnica de Catalunya, Mediterranean Technology Park, 08860 Castelldefels (Barcelona), Spain,
	\\
$^2$Institute of Spectroscopy, Russian Academy of Sciences, Troitsk, Moscow Region, 142190, Russia
\\
$^3$Centro de Fisica Te\'orica e Computacional and Departamento de F\'isica, Faculdade de Ci\^encias,
Universidade de Lisboa, Portugal
\\
$^4$Department of Physical Electronics, School of Electrical Engineering, Faculty of Engineering, Tel Aviv University, Tel Aviv 69978, Israel}




\begin{abstract}
We report on new types of two-component one-dimensional dark solitons (DSs) in a model of a dual-core waveguide with normal group-velocity dispersion and Kerr nonlinearity in both cores, the coupling between which is dispersive too. In the presence of the dispersive coupling, quiescent DSs supported by the zero-frequency background are always gray, being stable with the out-of-phase background, i.e., for opposite signs of the fields in the cores. On the contrary, the background with a nonzero frequency supports quiescent black solitons which may be stable for both out- and in-phase backgrounds, if the dispersive coupling is sufficiently strong. Only DSs supported by the out-of-phase background admit an extension to the case of nonzero phase mismatch between the cores.
\end{abstract}

\maketitle

Dark solitons (DSs) are fundamental modes in media where signs of nonlinear and dispersive terms in the nonlinear Schr\"{o}dinger (NLS)
equation, governing evolution of excitations, are opposite.
DSs were predicted in~Ref. \cite{Tsuzuki} in the context of the
mean-field theory of Bose-Einstein condensates (BECs), and obtained in the
framework of the inverse scattering method in~Ref. \cite{Zakharov}.
Experimentally, DSs were created in various physical systems, including optical fibers \cite{exp-fiber,exp-fiber2},
and BEC \cite{exp-BEC}.

DSs exist also in 
coupled NLS equations, including models of
dual-core optical couplers~\cite{AKA}. In the latter context, the inter-core
coupling may be dispersive, like the cores themselves \cite{Chiang1,Chiang2}%
, the effect that was observed experimentally~\cite{experiment}. The dispersive
coupling introduces new physics, 
as it links the temporal
structure of optical pulses with their energy distribution between the
cores, similar to the coupling of the translational and spinor degrees of
freedom in 
 spin-orbit-coupled (SOC)\ BECs~\cite{Spielman}.
In particular, the interplay of the SOC with the cubic self-attraction of
the condensate allows to suppress collapse and leads to formation of stable two-dimensional
(2D) bright solitons in the free space \cite{HS}. A similar mechanism produces stable families of spatiotemporal
optical bright solitons in a dual-core planar waveguides with the Kerr
self-focusing acting in each core \cite{we}. SOC in self-repulsive BECs in optical lattices supports vortex and half-vortex solitons~\cite{LKK}.
Similarity of mathematical description of evolution of SOC BEC and light propagation in waveguides with
dispersive coupling may allow transfer of many interesting concepts from the field of optics
to matter wave systems, and vice versa.

DSs in dispersively coupled waveguides were considered recently, mainly from
the perspective of the design of switching devices \cite{Sarma,GMU}. DSs
were also studied in SOC-BEC models~\cite{ASKFS,CSZQX}, where, unlike in the
optical setting, a trap potential is an inherent part of the physically
relevant model. In the presence of the trapping potential
DSs bifurcate from the first excited
state of the potential in the linear limit \cite{VK,Frantz}. The respective
quiescent DSs represent nonlinear modes with zero intensity at the center
\cite{ZDKM} (black solitons, BSs).

In this Letter, we introduce an
essentially new DS species in the dual-core waveguides, for which the
presence of the dispersive coupling is a {\em necessary} condition, i.e., they
disappear or become completely unstable when the dispersive part of the
coupling vanishes. We also for the first time illustrate nontrivial shape transformations of DSs due to considerable phase mismatch between the cores.
\begin{figure*}[ht!]
	\centering
	\includegraphics[width=\textwidth]{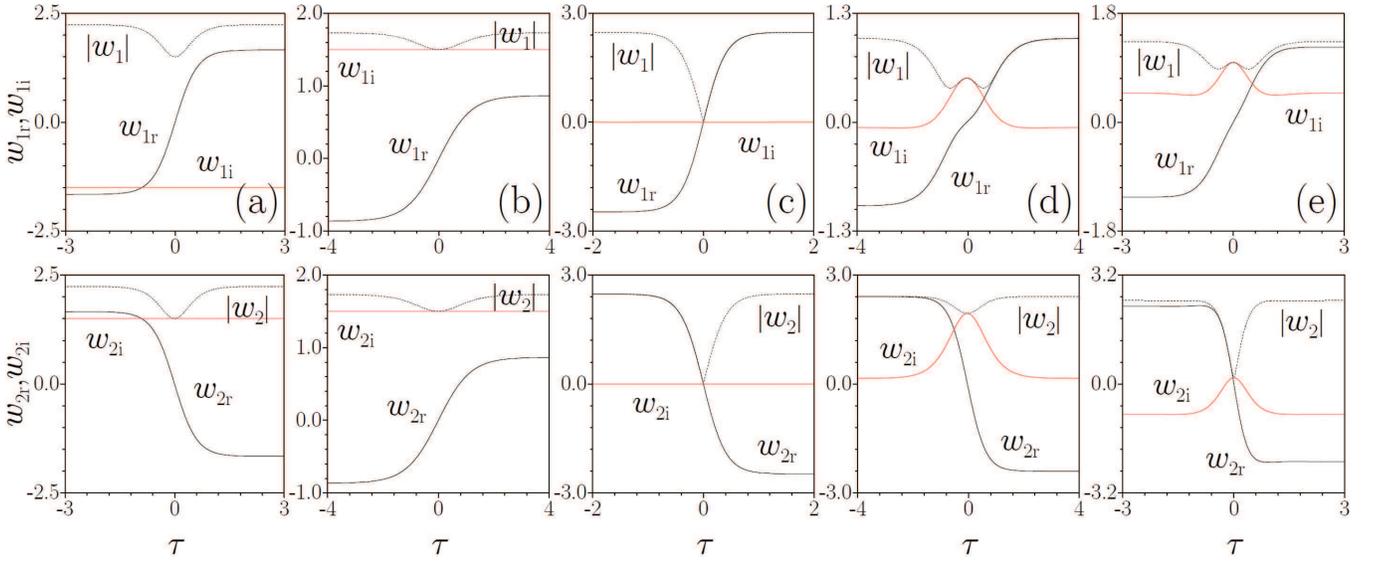}
	\caption{(Color online) Field profiles in the first and second (top and
		bottom rows) components for (a) out-of-phase GS, (b) in-phase GS, and (c)
		out-of-phase BS at $b=4$, $\protect\delta =1.5$, $\protect\beta =0$, as well
		as for (d) out-of-phase GS and (e) out-of-phase ``nearly-black" soliton
		with $b=2$, $\protect\delta =1$, $\protect\beta =3.4$
		(the latter one is a DS which would be black in the case $\protect\beta =0$%
		). Background oscillations in the BSs are not shown, as they may be removed
		by transformation $q_{1,2}\rightarrow q_{1,2}\exp (\pm i\protect\delta
		\protect\tau )$ in Eqs.~(\protect\ref{coupled_NLS}), see Eq. (\protect\ref%
		{black}). Here and in other figures $\kappa=1$.}
	\label{fig:one}
\end{figure*}

We consider a system of coupled NLS
equations for scaled field amplitudes $q_{1,2}$ in the two cores with the
normal intrinsic group-velocity dispersion and self-focusing Kerr
nonlinearity:
\begin{eqnarray}
\label{coupled_NLS}
\begin{array}{l}
\displaystyle{ i\frac{\partial q_{1}}{\partial z}-\frac{1}{2}\frac{\partial ^{2}q_{1}}{%
\partial \tau ^{2}}+i\delta \frac{\partial q_{2}}{\partial \tau }+\kappa
q_{2}+\beta q_{1}+\left\vert q_{1}\right\vert ^{2}q_{1} = 0},
\\[1mm]
\displaystyle{ i\frac{\partial q_{2}}{\partial z}-\frac{1}{2}\frac{\partial ^{2}q_{2}}{%
\partial \tau ^{2}}+i\delta \frac{\partial q_{1}}{\partial \tau }+\kappa
q_{1}-\beta q_{2}+\left\vert q_{2}\right\vert ^{2}q_{2} = 0}.
\end{array}
\end{eqnarray}%
Here $z$ and $\tau $ are the propagation coordinate and reduced time,
respectively \cite{Agrawal}, $\kappa >0$ is the linear coupling coefficient,
$\delta $ is the coupling-dispersion strength, and $\beta $ accounts for the
possible 
phase-velocity mismatch between the cores \cite{Dave}. In the 
most 
general form, we are interested in DSs
satisfying boundary conditions ($j=1,2$)
\begin{equation}
\lim_{t\rightarrow \pm \infty }q_{j}=Q_{j}\exp \left\{ i\left[ bz+\omega
\tau -(-1)^{j}\phi /2\pm \chi \right] \right\} ,
\label{boundary}
\end{equation}%
where $Q_{1,2}$ are the background amplitudes in the cores, $%
\phi $ is the phase difference between them, $b$ and $\omega $ are the
propagation constant and frequency, and $\chi $ is a phase shift across the
DS.

In this work, we focus on the settings with the in-phase ($\phi =0$) and
out-of-phase ($\phi =\pi $) backgrounds. Accordingly, we define $\sigma
\equiv \exp \left( i\phi \right) =\pm 1$. Then, the substitution of the ansatz (%
\ref{boundary}) in Eqs. (\ref{coupled_NLS}) connects the propagation
constant and phase-velocity mismatch with the asymptotic amplitudes of the background at $t\rightarrow \pm \infty$:
\begin{equation}
\label{Q}
b =\frac{(Q_{1}^{2}+ Q_{2}^{2})(\sigma \kappa+ Q_1Q_2)} {2Q_1Q_2},
 \quad
\beta  =\frac{(Q_{1}^{2}- Q_{2}^{2})(\sigma \kappa- Q_1Q_2)} {2Q_1Q_2}
\end{equation}

We start the analysis by producing exact DS solutions for $\beta =0$ (no
inter-core mismatch) and zero frequency, $\omega =0$. Then, in-phase ($%
\sigma =1$) and out-of-phase ($\sigma =-1$) DSs, which, generally speaking,
may move with nonzero velocity $v$, are obtained as $q_{1,2}=e^{ibz}w_{1,2}$%
, where
\begin{equation}
\begin{array}{c}
w_{1}=\sigma w_{2}=\left[ i(\sigma \delta -v)+\alpha _{\mathrm{g}}\tanh
(\alpha _{\mathrm{g}}(\tau -vz))\right] ,
\\
a_{\mathrm{g}}^{2}\equiv b-\sigma \kappa -(\sigma \delta -v)^{2}
\end{array}
\label{alpha}
\end{equation}%
(subscript $\mathrm{g}$ stands for ``gray", see below).
These DSs exist above the propagation-constant cutoff corresponding to $a_{g}^{2}\geq 0$, i.e., $b\geq (\sigma \delta -v)^{2}+\sigma \kappa $.  
A representative feature of the DSs in the system with the dispersive
coupling ($\delta \neq 0$) is that the quiescent solitons, with $v=0$,
unlike the conventional BSs, keep finite ``grayness", which
may be characterized by the intensity of each component at the soliton's
center, $r_{\mathrm{g}}=\delta ^{2}$, as it follows from Eq.~(\ref{alpha}).
Therefore, the quiescent DSs with zero frequency $\omega$ are called gray solitons (GSs) below. Examples
of complex profiles, $w_{1,2}\equiv w_{1\mathrm{r},2\mathrm{r}}(\tau )+iw_{1%
\mathrm{i},2\mathrm{i}}(\tau )$, of the in-phase and out-of-phase GSs are displayed
in Figs.~\ref{fig:one}(a) and (b), respectively.

A second representative
family of solutions of Eqs. (\ref{coupled_NLS}) at $\beta=0$ corresponds to solitons with nonzero internal frequency $\omega$.
Such solitons, satisfying boundary conditions (\ref{boundary}), can be found in the form $q_{1,2}=e^{ibz+i \omega \tau}w_{1,2}$, where 
$\omega=\delta$ and $\omega=-\delta$ respectively for the families with in-phase and out-of-phase first and second components. Such solutions have the form:
\begin{gather}
w_{1}=\sigma w_{2}=\alpha _{\mathrm{b}}\tanh (\alpha _{\mathrm{b}}\tau
),\quad
a_{\mathrm{b}}^{2}=b-\sigma \kappa +\delta ^{2}/2.
\label{black}
\end{gather}%
with $\sigma$ defined above. The intensity distribution of such solitons resembles that in classical black states with zero intensity at $\tau=0$
(hence we call this family BSs and use subscript ``b" to distinguish it from the GS family), but at the same time they reside on the background wave with nonzero phase tilt $\sigma \delta \tau$.
An example of an out-of-phase BS is shown in Fig.~\ref{fig:one}(c).

\begin{figure}[h]
\centering
\includegraphics[width=\linewidth]{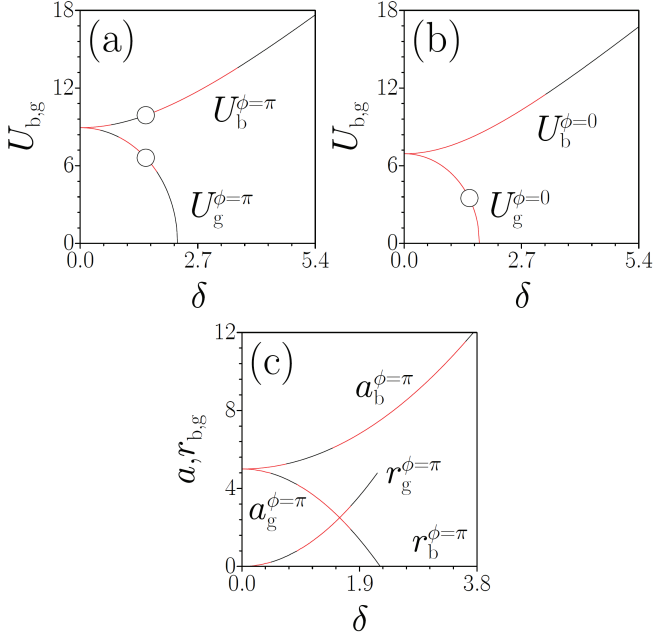} 
\caption{(Color online) The renormalized energy flow of out-of-phase [marked
``$\protect\phi =\protect\pi $" in panel (a)] and in-phase
[``$\protect\phi =0$" in (b)] BSs and GSs (subscripts
``b" and ``g", respectively), {\em vs} the
coupling-dispersion coefficient, $\protect\delta $. (c) The amplitude, $a_{%
\mathrm{b,d}}$, and grayness, $r_{\mathrm{b,d}}$, of the out-of-phase BSs
and GSs {\em vs} $\protect\delta $. In all cases $b=4$, $\protect%
\beta =0$. Black (red) curves denote stable (unstable) branches.
Circles in (a) and (b) correspond to the GSs and BS displayed
in Figs.~\protect\ref{fig:one}(a,c) and (b).}
\label{fig:two}
\end{figure}

BS and GS families are represented in Fig. \ref{fig:two} by the 
renormalized energy flow, defined as $U=U_{1}+U_{2}$, with $%
U_{1,2}\equiv \int_{-\infty }^{+\infty }\left( |Q_{1,2}|^{2}-\left\vert
w_{1,2}(\tau )\right\vert ^{2}\right) d\tau $. For $\beta =0$, Eqs. (\ref{alpha}) and (\ref{black}) yield $U_{\mathrm{g,b}}=4a_{\mathrm{g,b}}$. The
figure also demonstrates alternation of stable (black) and unstable (red) branches within each subfamily.
The instability may stem from the modulational instability (MI) of the background, and from the instability of the localized DS's core. The MI is amenable to
analytical investigation at $\beta=0$. In particular, 
one can show that 
at $\omega =0$ the out-of-phase background is always stable, while in-phase background is unstable.
At $\omega\neq 0$ the out-of-phase background is also stable, while its
in-phase counterpart is unstable at $\delta ^{2}<\kappa$, and becomes stable for
sufficiently strong coupling dispersion, $\delta ^{2}>\kappa$. This result is
explained by the fact that substitution $q_{1,2}\equiv \tilde{q}%
_{1,2}\exp \left( i\delta \tau \right) $, see Eq. (\ref{black}), transforms
Eqs. (\ref{coupled_NLS}) into similar equations
with coupling constant $\tilde{\kappa}\equiv \kappa-\delta ^{2}$. The sign flip
of $\tilde{\kappa}$ at $\delta ^{2}>\kappa$ is tantamount to replacing the
in-phase configuration by an out-of-phase one, which leads to the
stabilization.

We now turn to the effect of the phase-velocity mismatch, $\beta $, on the
existence and stability of two above-mentioned families of DSs. In this case, Eqs.~(\ref{Q}) predicts
that background amplitudes for two components are different. For small $\beta\ll 1$, using (\ref{alpha}), one obtains $Q_{j}=(b-\sigma)^{1/2}-(-1)^j\epsilon+{\cal O}(\epsilon^2)$, where $\epsilon\equiv\beta (b-\sigma\kappa)^{1/2}/[2(2\kappa\sigma- b)]$. Considering the simplest case of $\delta=0$ (generalization for $\delta>0$ is straightforward) one can rewrite Eqs.~(\ref{coupled_NLS})
in terms of the normalized fields $p_{j}=w_{j}/Q_{j}$ ($j=1,2$):
\begin{equation}
(1/2)p_{j,\tau\tau} + \kappa(Q_{3-j}/Q_{j})%
(p_{j}-p_{3-j})-Q_j(|p_{j}|^{2}-1)p_{j}=0.
 \label{dark_p}
\end{equation}%
We look for solutions of these equations in the form of $p_{j}=p-\epsilon v_j$, where $p$ is the BS solution of equation $p_{\tau\tau}-2(|p|^{2}-1)p=0$ and $v_{1,2}(\tau )\rightarrow 0$ at $\tau \to \pm \infty $. One can show that $v_1=- v_2$ and these functions are real, while $v_1$ satisfies linear equation $v_{1,\tau \tau }+2\left(2+\sigma-3p^{2}\right) v_1=4p\left( 1-p^{2}\right) $.  The asymptotic form of this equation at $\tau \rightarrow \infty $ is given by $v_{1,\tau \tau }=0$ for $\sigma=1$, and $v_{1,\tau \tau }-4v_1=0$ for $\sigma=-1$. This shows that for $\beta>0$ there may be no exponentially decaying solutions, i.e., no extension to $\beta\neq 0$ is possible, for the in-phase solitons with $\sigma=1$, while out-of-phase solitons with $\sigma=-1$ may exist.

Systematic results for $\beta \neq 0$ were produced in a numerical
form. In accordance with the above conclusion, DSs exist solely with the
out-of phase background, see examples of profiles of GS and BS in Figs.~\ref%
{fig:one} (d,e), respectively. At large $\beta $, the DSs of both types feature a double-well structure
of the first component (weakly pronounced in these examples), and the local
intensity never vanishes (hence, the notation BS serves here only for stressing that this family resides on the background wave $exp(-i \delta \tau)$ and that it becomes black at $\beta=0$). Further increase of $\beta $ reveals new shapes of
DSs, with the first component featuring an elevation (rather than
depression) against the background.

Properties of the DSs for $\beta \neq 0$ are summarized in Fig.~\ref%
{fig:three}. Note a non-monotonous dependence of the renormalized energy
flow on $\beta $ for both GS and BS families. The existence of the GS family is limited to a finite
domain, $\beta <\beta _{\mathrm{cr}}$. The respective derivative $dU/d\beta $ diverges at $\beta
=\beta _{\mathrm{cr}}$. The existence of this family is also restricted to
finite values of the coupling dispersion, $\delta $. On the other hand, no
such bounds were found for the branches originating from the BS at $\beta =0$.
\begin{figure}[tbph]
\centering
\includegraphics[width=\linewidth]{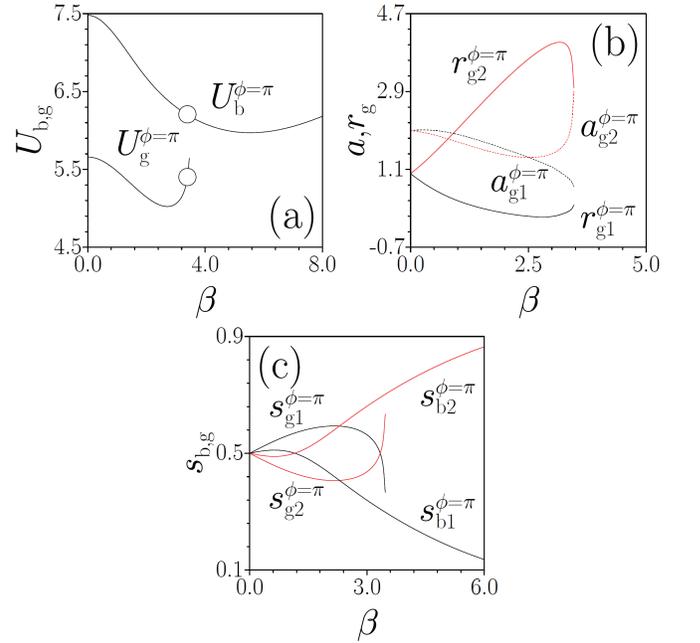} 
\caption{(Color online) The renormalized energy flow of the out-of-phase
near-BSs (DSs, which become black at $\protect\beta =0$) 
and GSs (a), and the energy-flow shares in the two components, $s_{1,2}$, (c), {\em vs} the
inter-core mismatch $\protect\beta $ at $b=2$, $\protect\delta =1$. The
dashed and solid curves in (b) show, respectively, the
amplitude ($a$) and grayness ($r$) of GSs {\em vs} $\beta$. All families depicted in the figure are stable. Circles in
(a) correspond to the solitons in Fig.~\protect\ref{fig:one}(d),(e).}
\label{fig:three}
\end{figure}

In Fig.~\ref{fig:three}(b) we observe that $\beta \neq 0$ results in
differences of the soliton amplitudes and grayness in the components, in
addition to the difference in the background amplitudes. The most pronounced
difference is observed for the grayness, with a relatively weak amplitude
mismatch, almost in the whole existence domain, except for a narrow region
near $\beta _{\mathrm{cr}}$, where the amplitude of the first (second)
component abruptly decays (grows). Shares of the renormalized energy in the DS components, $s_{j}\equiv U_{j}/U$%
, which are displayed in Fig.~\ref{fig:three}(c), exhibit a non-monotonous
behavior: the share of the first component in GS initially grows with $\beta $, but near $%
\beta _{\mathrm{cr}}$ it suddenly decreases and renormalized power of second components starts to dominate. A
smoother non-monotonous dependence $s_{1,2}(\beta )$ is also featured by the
near-BS branches that do not have any cutoff in $\beta$.

\begin{figure}[t]
	\centering
	\includegraphics[width=\linewidth]{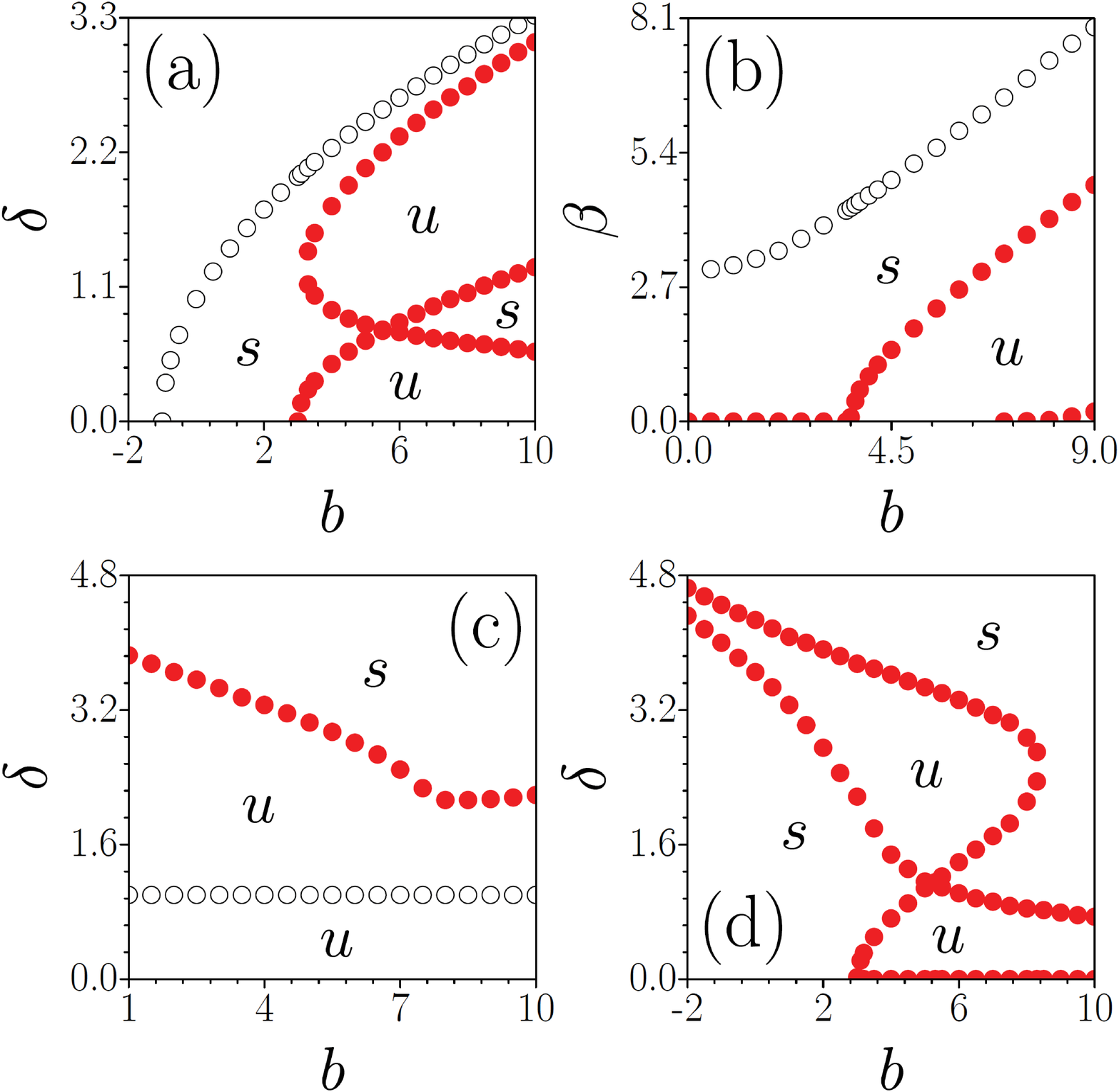} 
	\caption{(Color online) Existence and stability domains for out-of-phase GSs
		in the plane of $(b,\protect\delta )$ at $\protect\beta =0$ (a), and in the
		plane of $(b,\protect\delta )$ at $\protect\delta =1$ (b); for the in-phase
		BSs in the plane of $(b,\protect\delta )$ at $\protect\beta =0$ (c), and for
		the out-of-phase BSs in the plane of $(b,\protect\delta )$ at $\protect\beta %
		=0$ (d). Stability and instability areas are marked by ``s"
		and ``u", respectively. In (a) and (b), lines of open
		circles show the upper boundary of the existence domain. In (c) and (d)
		there is no upper existence boundary, while the horizontal line of open
		circles in (c) indicates the border where the background for the in-phase
		BSs becomes modulationally stable, $\protect\delta ^{2}=1$, see the text.
		In-phase GSs are supported by the modulationally unstable background,
		therefore they are not shown.}
	\label{fig:four}
\end{figure}

Stability of all types of DSs described here was thoroughly analyzed by means of computing the respective
eigenvalues, using equations for infinitesimal perturbations linearized
around the stationary DSs. Results are summarized in Fig.~\ref{fig:four}.
For $\beta =0$, the out-of-phase gray solitons ($\sigma =-1$) exist below the
line of empty circles, given by $\delta ^{2}=b- \sigma \kappa$, as
obtained from $a_{\mathrm{g}}=0$, see Eq. (\ref{alpha}) with $v=0$. GSs
are stable for relatively small $b$. A complex instability domain appears as
$b$ increases. For large propagation constant values GSs are unstable for small $\delta$, while the
increase of the coupling dispersion eventually stabilizes them. The dependence of stability domains for the out-of-phase GSs on $\beta $ at fixed $\delta =1$ is shown
in Fig.~\ref{fig:four}(b). For small $b$, out-of-phase GSs are stable in their
existence domain, which is located below the line of empty circles. For
larger $b$, they are stable close to the upper boundary of their existence domain. In-phase GSs are always unstable and their existence domain is not shown here.

Stability analysis for the in-phase BSs on the plane $(b,\delta)$ is presented in Fig.~%
\ref{fig:four}(c), where the stabilizing role of the dispersive coupling is
evident. Note that background of such solitons is modulationally unstable below line of open circles. Above this line there is a domain of instability of localized soliton core, but at large $\delta$ values such solitons become completely stable. Finally, Fig.~\ref{fig:four}(d) shows rather complex stability and
instability domains for the out-of-phase BSs on the plane of parameters $(b,\delta)$. Here, increasing dispersion of coupling also stabilizes solitons. The domains of stability for BSs on the $(b,\beta)$ plane are not shown, but at $\delta=1$ such solitons are always stable at $b<4.75$.

In conclusion, we have reported new types of two-component DSs (dark
solitons) in the model of two-core waveguide with normal
dispersion in both cores and dispersive coupling between them. The
inter-core phase-velocity mismatch was included too. Due to the presence of
the coupling dispersion, zero-frequency background always supports GSs (gray
solitons), which are stable only in the case of the out-of-phase background,
with opposite signs of the fields in the two cores. On the other hand, the
background with a nonzero frequency supports BSs (black solitons), which may
be stable for the in-phase background too, provided that the coupling
dispersion is strong enough. Solely the out-of-phase background admits the
extension of DSs to the system with the phase-velocity mismatch between the
cores.


VVK was funded by Funda\c{c}\~ao para a Ci\^encia e
Technologia (Portugal), grant UID/FIS/00618/2013.


\end{document}